\documentclass[12pt]{article}

\usepackage{graphicx}

\newcommand{\lsim}{\raisebox{-0.13cm}{~\shortstack{$<$ \\[-0.07cm] $\sim$}}~}

\begin{document}

\mbox{ } \\[-1cm]
\mbox{ }\hfill hep-ph/0605273
\bigskip
\thispagestyle{empty}
\setcounter{page}{0}
\begin{center}
{\Large{\bf
 Addendum to ``Threshold corrections to $m_b$ and the         \\[0.7mm] 
 $\bar{b}b\to H^0_i$ production in
 CP-violating SUSY scenarios''                                \\[5mm]
}}
{\large
 Francesca Borzumati$^{1,2}$ and Jae Sik Lee$^3$}
\end{center}
\begin{center}
$^1$
{\it International Center for Theoretical Physics, 
     Trieste, Italy}                        
\\[1mm] 
$^2$
{\it Scuola Internazionale Superiore di Studi Avanzati, 
     Trieste, Italy}
\\[1mm]
$^3${\it 
 Centre for Theoretical Physics, 
Seoul National University, Seoul, 
 Korea}
\end{center}
\vskip 1.5cm

\begin{abstract}
In hep-ph/0401024:                       
``Threshold corrections to $m_b$ and the $\bar{b}b\to H^0_i$ 
production in CP-violating SUSY scenarios'', we have pointed out 
that the production cross sections of the three neutral Higgs
bosons through $\bar{b}b$ fusion can deviate substantially from 
those obtained in CP conserving scenarios, thanks to the 
nontrivial role played by the threshold corrections to $m_b$
combined with the CP-violating mixing in the neutral-Higgs-boson 
sector. 
The deviations are largest for values of the CP violating phases 
that maximize the mixing among at least two of the three 
neutral Higgs bosons.
We complement our previous work focussing explicitly on the values 
of masses and widths of the three neutral Higgs bosons in this 
region of parameter space. We then address the issue of whether 
the three different peaks in the invariant mass distribution of
the Higgs-decay products can be experimentally disentangled
at the LHC.   
\end{abstract}


\newpage 
\setcounter{page}{1}
\setlength{\parskip}{1.01ex} 

The production cross sections of the three neutral Higgs
bosons through $b$-quark fusion can deviate substantially from 
those obtained in CP conserving scenarios, thanks to the 
nontrivial role that the threshold corrections to 
$m_b$ can play in these scenarios~\cite{Borzumati:2004rd}. 
The largest deviations in the case of $H_1$ and $H_2$ 
are for values of $\Phi_{A\mu}$ 
around $100^\circ$, with a large enhancement for 
the production cross section of 
$H_1$, a large suppression for that of $H_2$. The former is due 
to the fact that the component of the field $a$ in $H_1$ around these
values of $\Phi_{A\mu}$ is large, while it is depleted by a similarly
large amount in $H_2$. 
The cross section for $H_3$ is also largely affected by the $m_b$ 
corrections, 
but this deviation 
is roughly independent of $\Phi_{A\mu}$.

\begin{figure}[!h]
\begin{center}
\includegraphics[width=5.1cm]{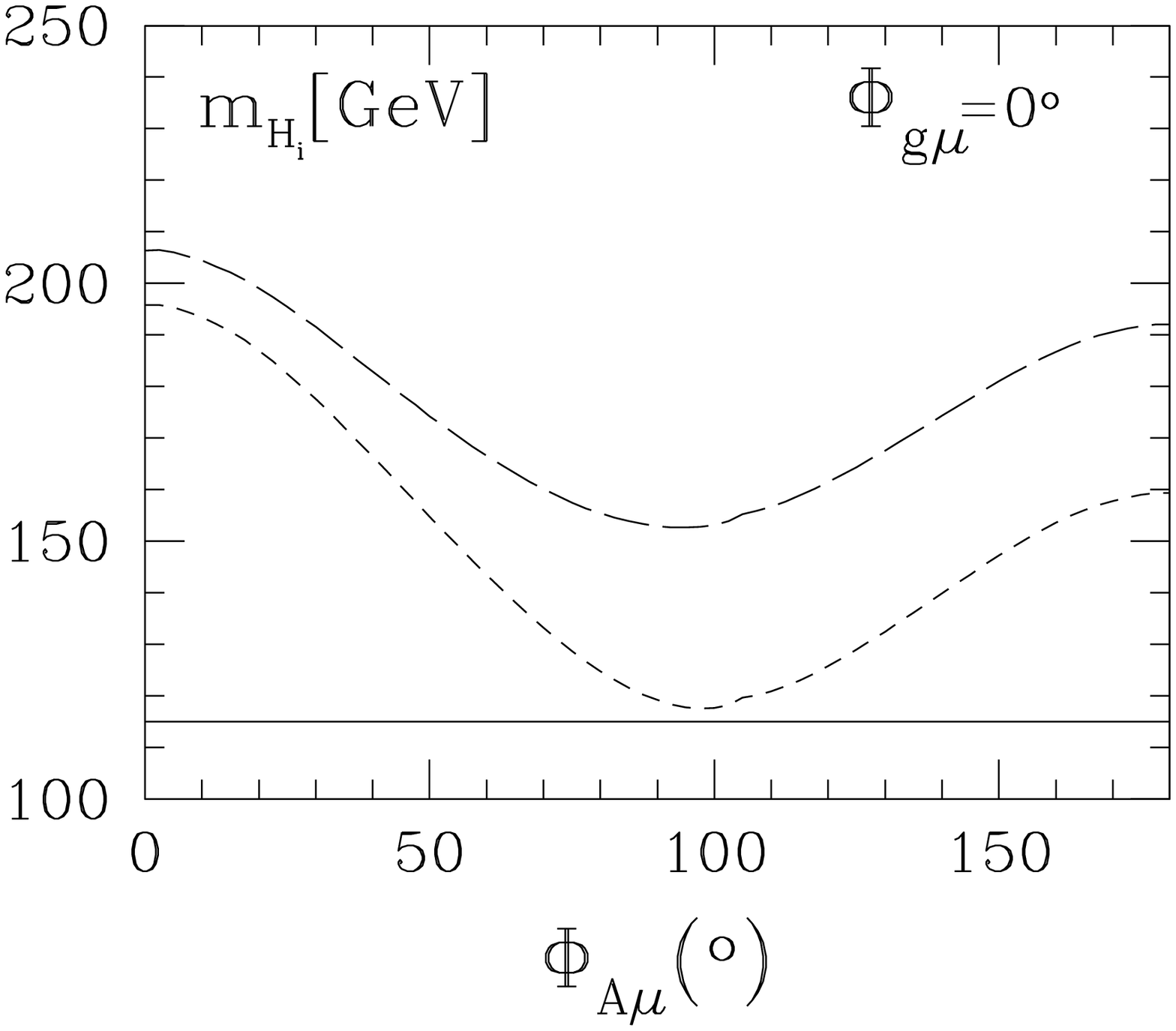}
\hspace{1mm}
\includegraphics[width=5.1cm]{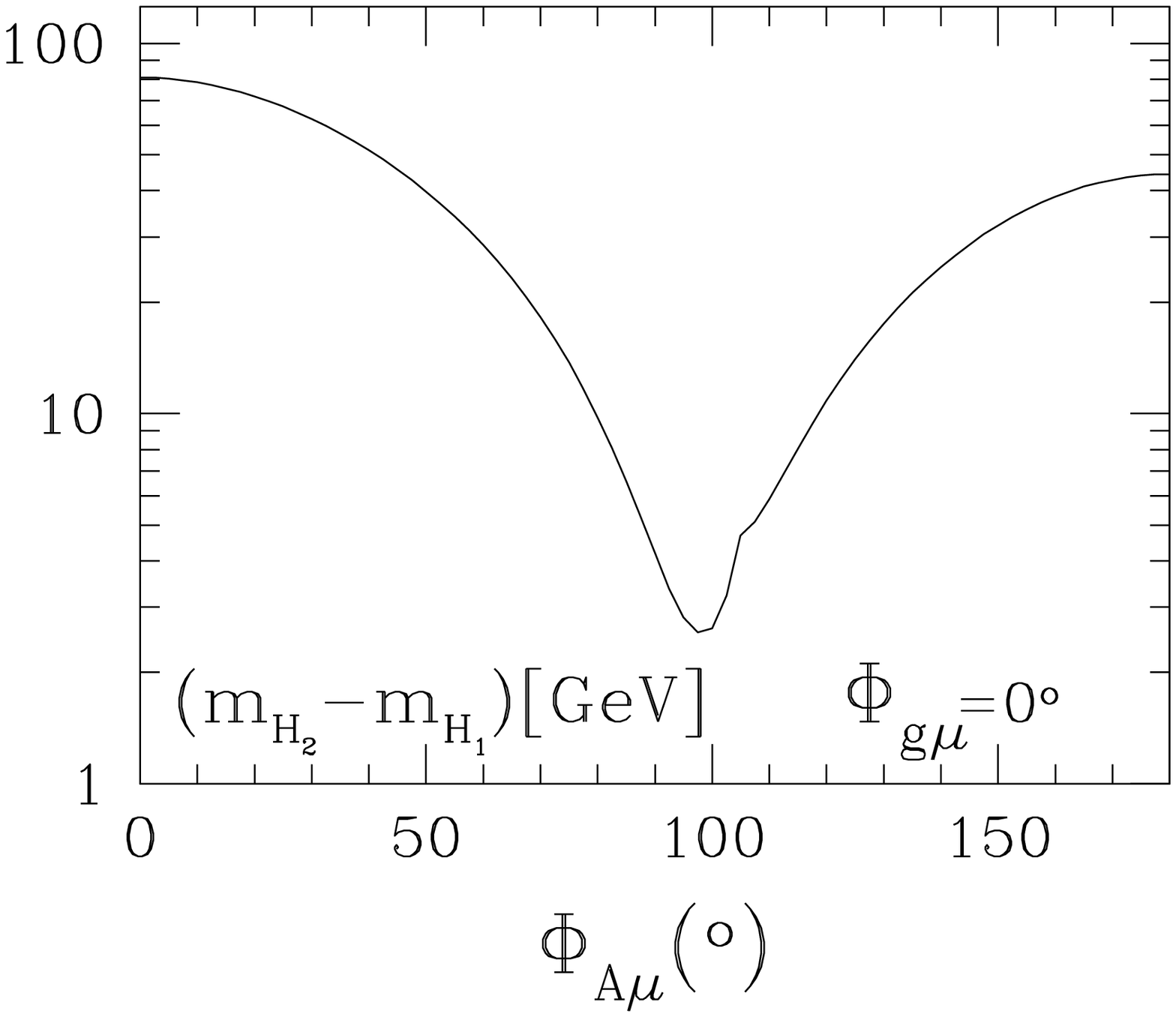}
\hspace{1mm}
\includegraphics[width=5.1cm]{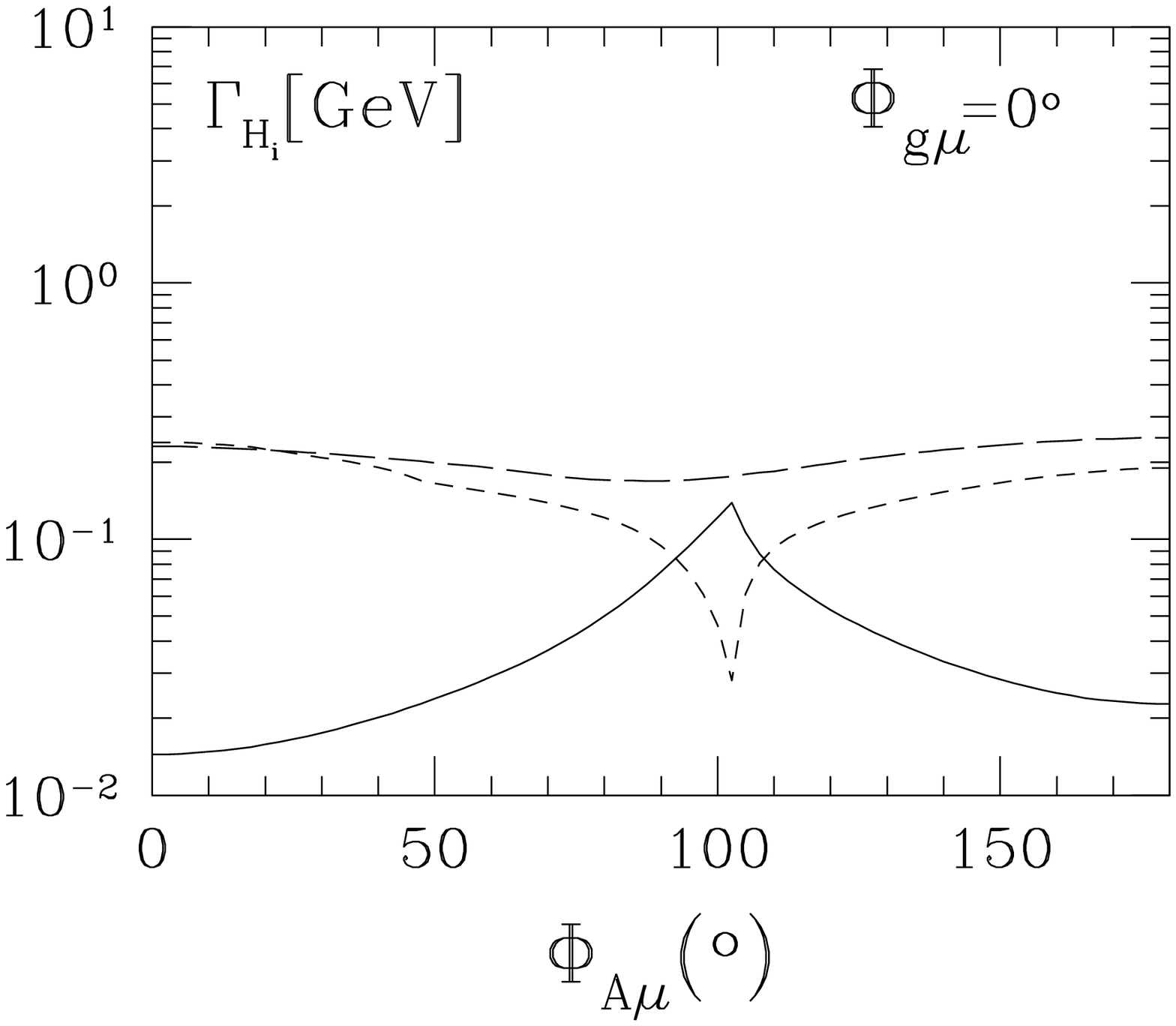}
\end{center}
\vspace{-0.4cm}
\begin{center}
\includegraphics[width=5.1cm]{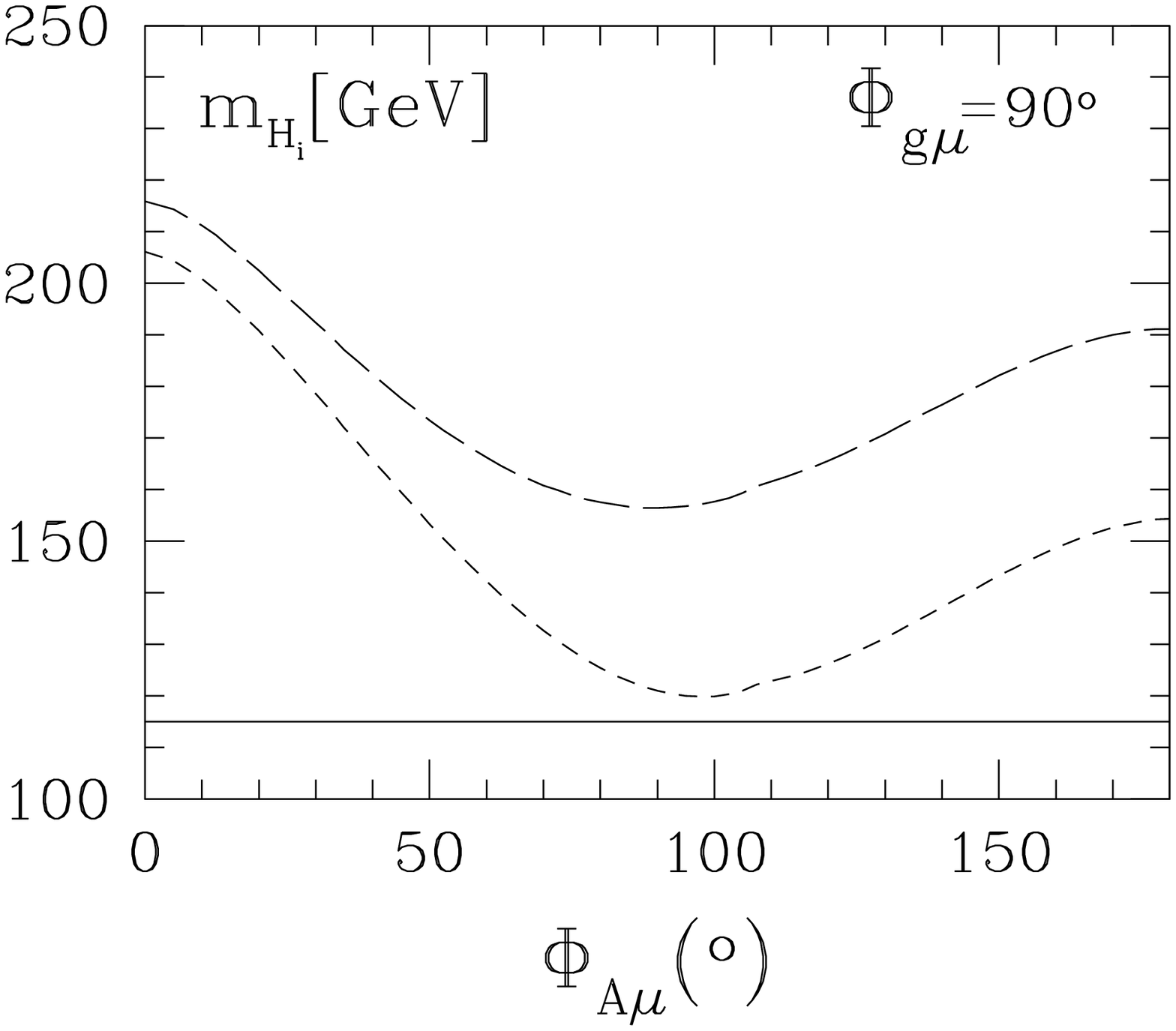}
\hspace{1mm}
\includegraphics[width=5.1cm]{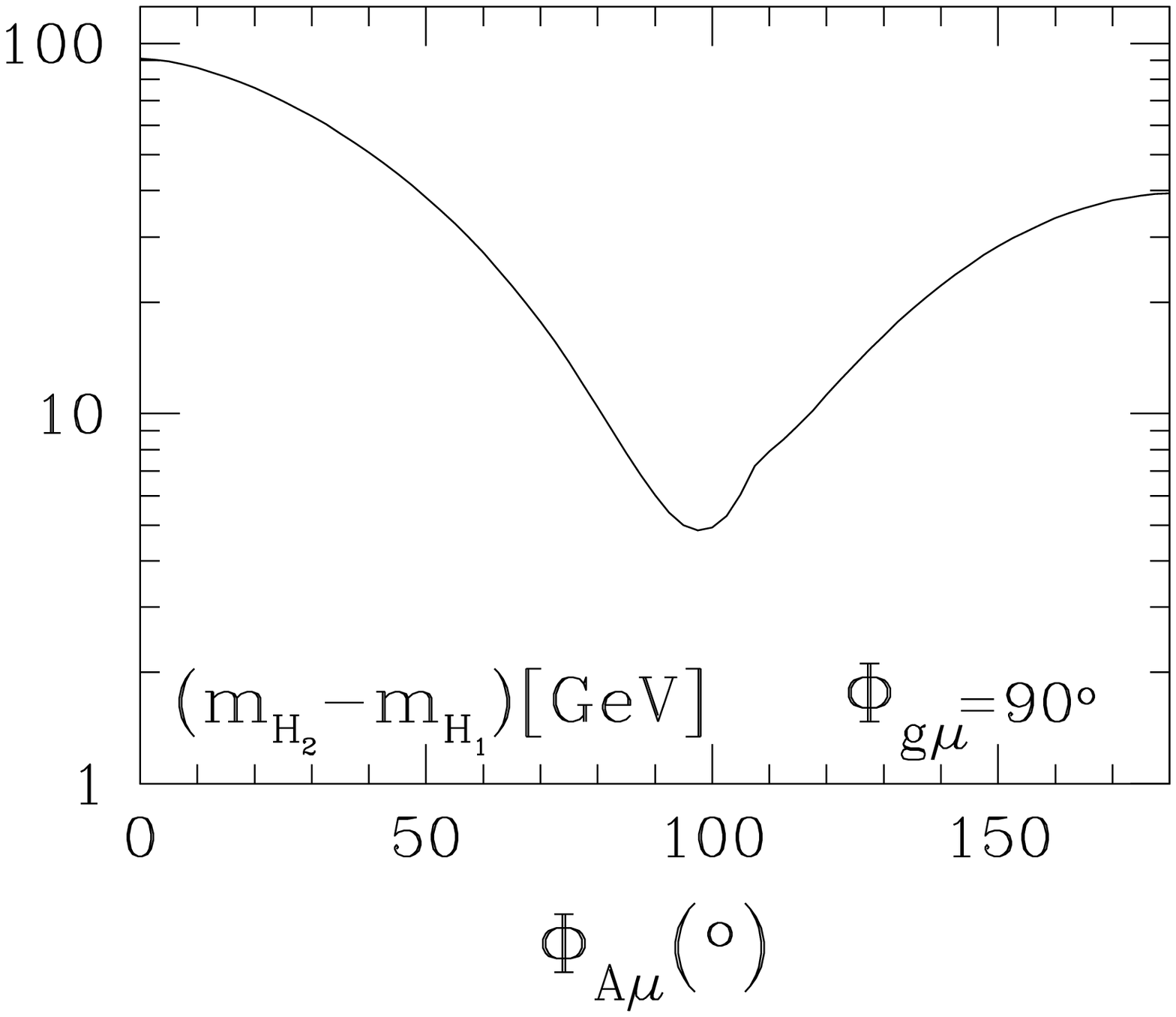}
\hspace{1mm}
\includegraphics[width=5.1cm]{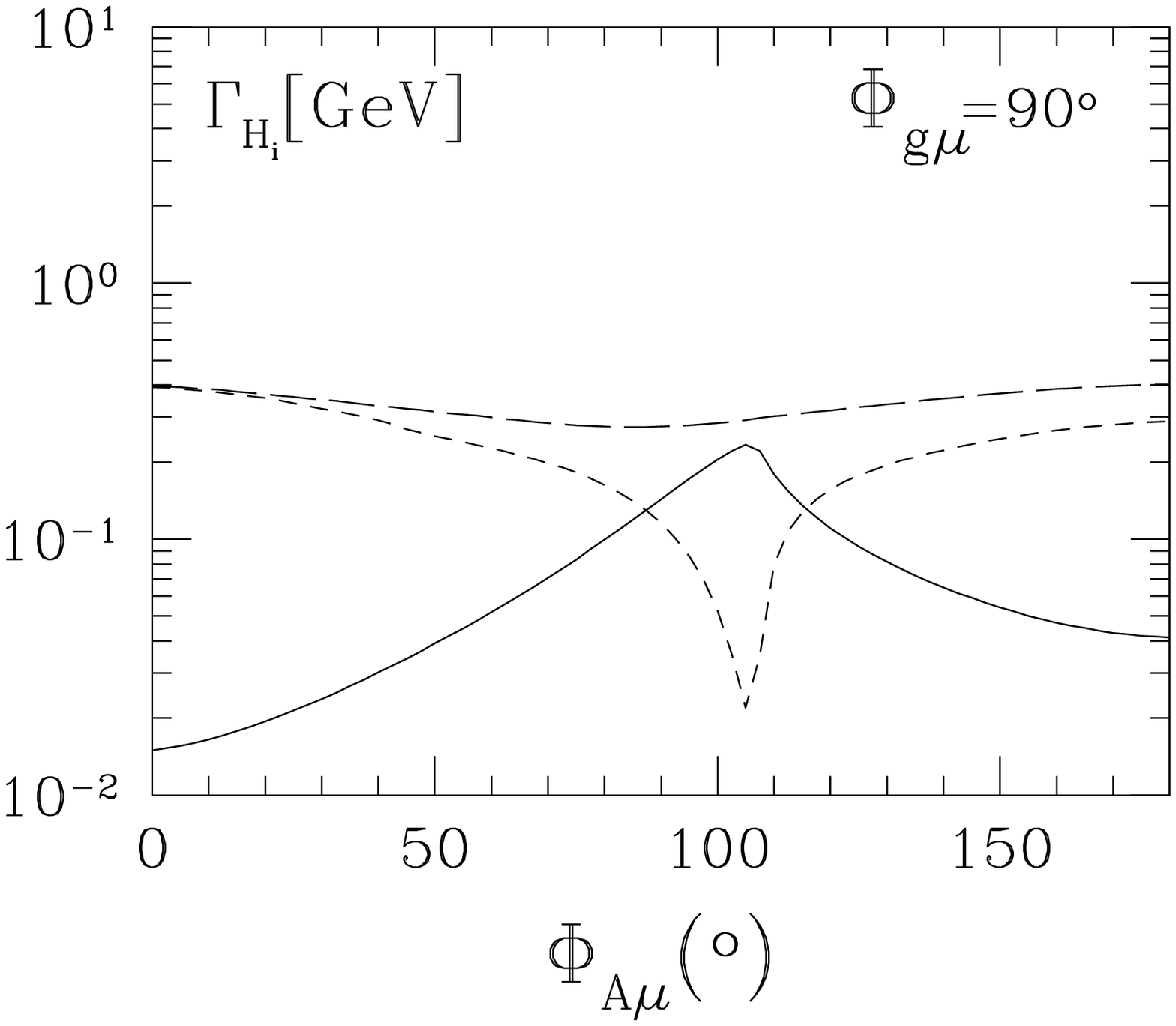}
\end{center}
\vspace{-0.4cm}
\begin{center}
\includegraphics[width=5.1cm]{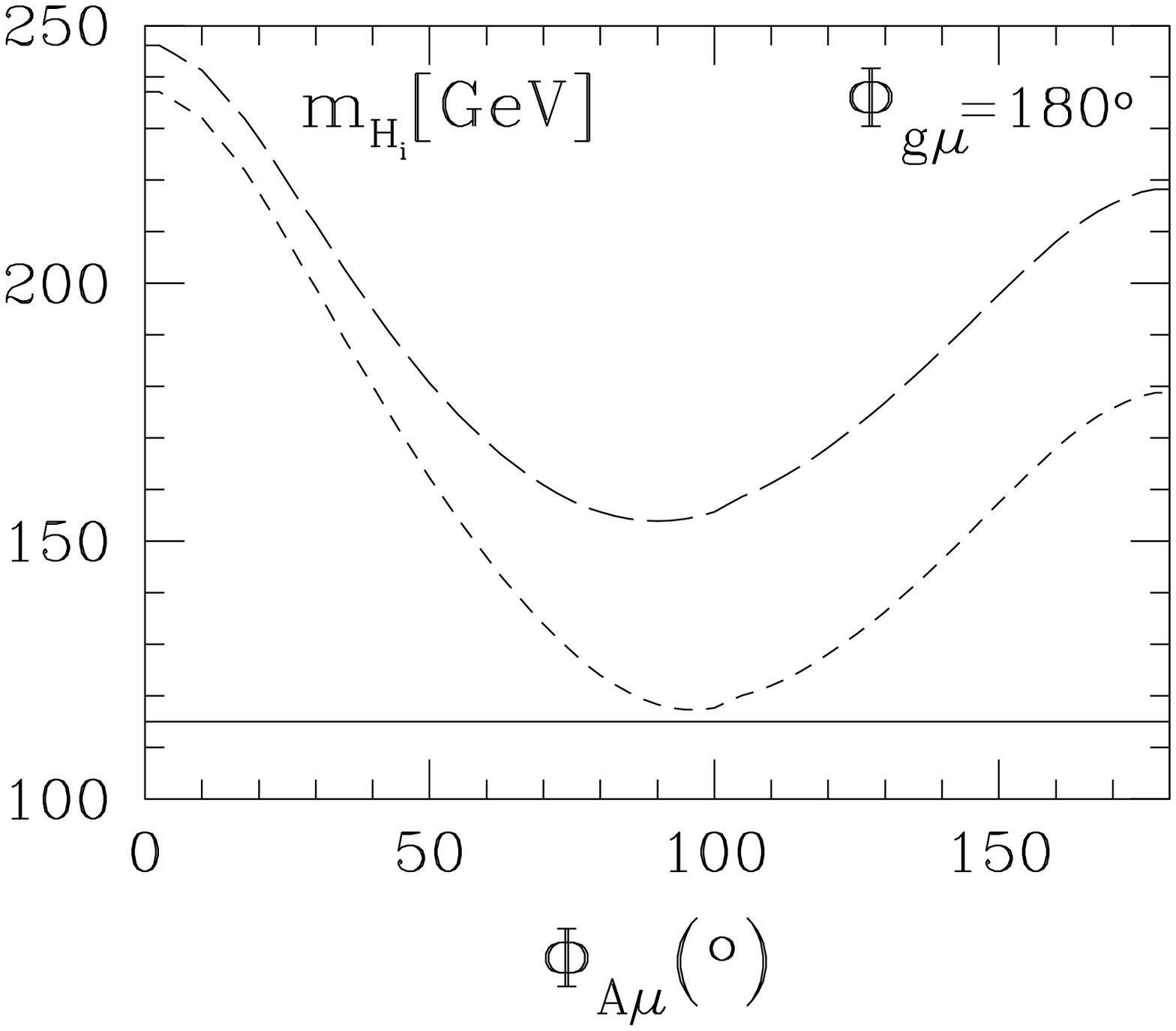}
\hspace{1mm}
\includegraphics[width=5.1cm]{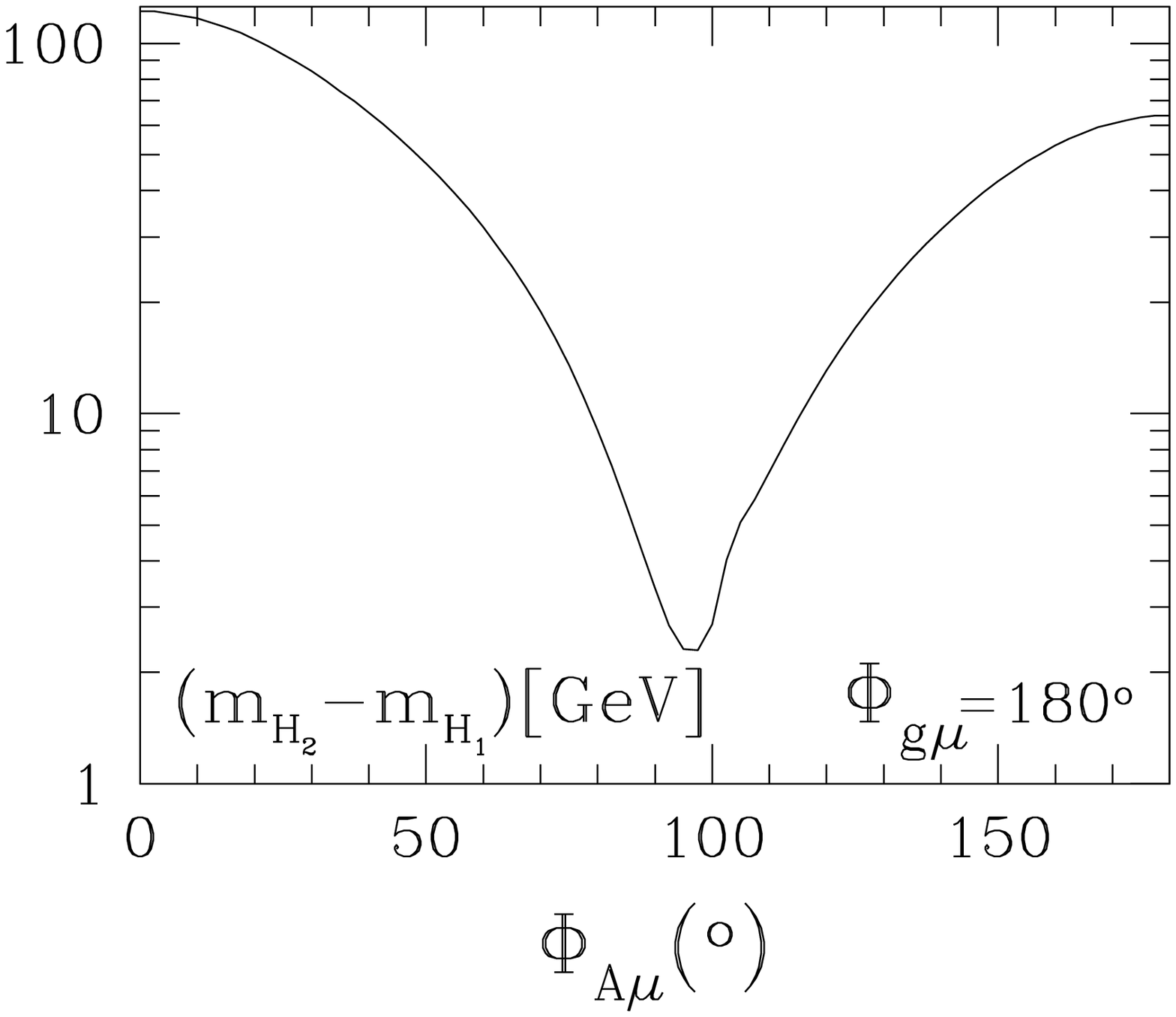}
\hspace{1mm}
\includegraphics[width=5.1cm]{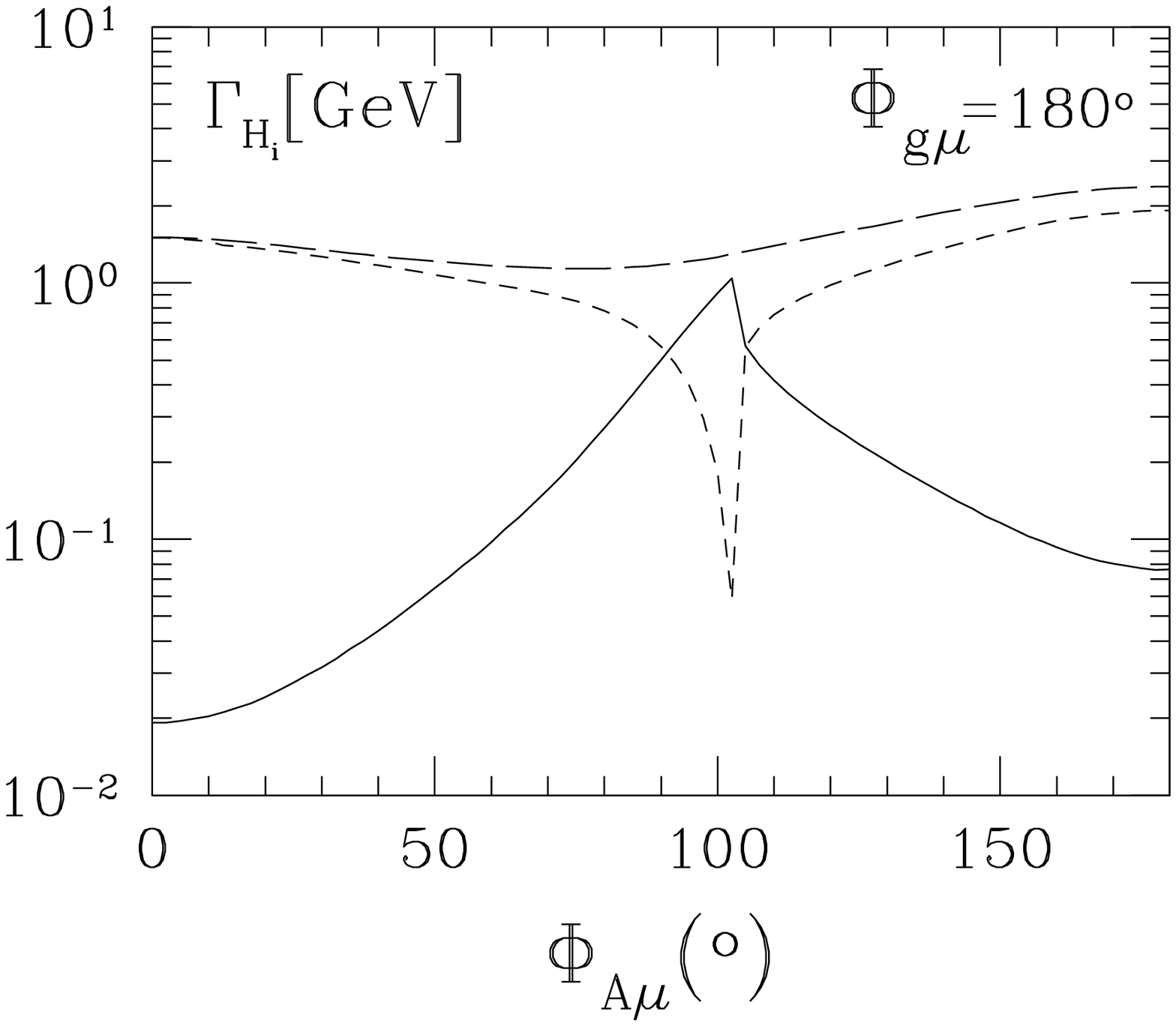}
\end{center}
\vspace{-0.4cm}
\caption{{\small \it The masses (left column) and 
 widths (right column) of the neutral Higgs
 bosons as functions of $\Phi_{A\mu}$ for the CPX spectrum
 specified in Ref.~\cite{Borzumati:2004rd} with
 $M_{\rm SUSY}=0.5\,$TeV and
 $\tan\beta=10$. Three values of $\Phi_{g\mu}$ are considered:
 $0^\circ$ (top row), $90^{\circ}$ (middle row) and 
 $180^\circ$ (bottom row). The solid lines are for
 $H_1$, the dashed ones for $H_2$, and the long-dashed ones for
 $H_3$. The central column is for the mass difference between $H_2$
 and $H_1$.}}
\label{fig:mass_width}
\end{figure}

In this region of large mixing the $H_1$ and $H_2$ bosons have 
very similar masses, as shown in the first column of 
Fig.~\ref{fig:mass_width}, for three different values of 
$\Phi_{g \mu}: 0^{\circ}, 90^{\circ}, 180^{\circ}$. In the 
first and the third column of this figure, the solid lines 
always represent $H_1$, the dashed lines $H_2$, the long-dashed 
ones $H_3$. 
At $\Phi_{A\mu} =100^{\circ}$, the $H_1$-$H_2$ mass difference 
is always below $5\,$GeV,
as shown by the second column of the same figure.  
As for the widths of these neutral Higgs bosons, $\Gamma_{H_1}$
is always about 10 times larger than $\Gamma_{H_2}$
at $\Phi_{A\mu} =100^{\circ}$. See the 
third column of Fig.~\ref{fig:mass_width}.
Notice that always at $\Phi_{A\mu} =100^{\circ}$, 
a factor of 10 is also the ratio 
of the production cross sections of $H_1$ and $H_2$, both
at the LHC and at the Tevatron.

Given the degeneracy between $H_1$ and $H_2$, it is legitimate
to worry whether a transition $H_1 \rightarrow H_2$ can 
occur during propagation and before decay, due to the 
off-diagonal absorbitive parts in the $3 \times 3$ matrix for the
neutral Higgs boson propagator considered in 
Ref.~\cite{Ellis:2004fs}. We have numerically checked the size of
these off-diagonal parts and found that in our specific case 
they are negligible. We have nevertheless included these terms
in our numerical calculations. Thus, 
near $\sqrt{\hat{s}}=m_{H_i}$, the partonic cross section 
for the $\bar{b} b$-fusion production of $H_i$ and their 
subsequent decays into a final state $f.s.$,   
$\hat{\sigma}(\bar{b} b \to H_i \to f.s.)$, 
hereafter denoted as $\hat{\sigma}^{f.s.}$,
is given, to a very good approximation, 
by the cross section with a single $\hat{s}$-channel resonance 
with mass $m_{H_i}$ and width $\Gamma_{H_i}$. 
Away from $\sqrt{\hat{s}}=m_{H_i}$,
all Higgs bosons $H_i$ contribute to the partonic cross section
for the production of the same final state $f.s.$.

The mass difference between $H_1$ and $H_2$ is, however, still 
small enough to question whether 
it is possible to
disentangle the two corresponding peaks in the 
invariant mass distributions of the 
$H_1$- and $H_2$-decay products.  
There is no similar problem for the 
$H_3$ eigenstate, that has a mass always larger than 
$\sim 160\,$GeV and therefore a splitting from $H_2$ always 
larger than $\sim 10\,$GeV. 
Since $\Gamma_{H_2\,,H_3} \lsim 2-3\,$GeV, 
we assume that this splitting can be experimentally
resolved. 
As already observed, on the contrary, in the case of $H_1$ and
$H_2$, the mass difference can be as small $2\,$GeV around
$\Phi_A=100^\circ$.
It will therefore be very challenging to disentangle
$H_2$ from $H_1$ experimentally.  
An analysis of these two Higgs bosons decay modes, of 
their differential cross section with respect to the 
invariant mass distribution of the decay products, and the
experimental resolution of these decays, can help in this sense.

For our discussion we shall concentrate on the issue of 
production and possible problem of detection of $H_1$ and
$H_2$ at the LHC only, where 
the best energy and momentum resolutions are for the 
Higgs-boson decays into muon and photon pairs. 
For these two decay modes, the invariant-mass resolutions
are, respectively, 
$\delta M_{\gamma\gamma}\sim 1\,$GeV and
$\delta M_{\mu\mu} \sim 3\,$GeV for a Higgs mass of
$\sim 100\,$GeV~\cite{ATLASTDR:1999fr}.

The $H_i$-differential production cross section through $b$-quark 
fusion with respect to the invariant mass distribution of the 
final state $f.s.$ ($ =\gamma \gamma$ or $\bar{\mu} \mu$) is 
\begin{equation}
\frac{{\rm d}\sigma^{f.s.}}{{\rm d} \sqrt{\hat{s}}} 
=
\frac{2}{\sqrt{\hat{s}}}\,\hat{\sigma}^{f.s.}(\tau)
\left(\!\tau\frac{\rm{d}{\cal L}^{\bar{b}b}}{{\rm d}\tau}\!\right)
=
\frac{2}{\sqrt{\hat{s}}}\,\hat{\sigma}^{f.s.}(\tau)
 \int_{\tau}^1 {\rm d}x
 \left[\frac{\tau}{x}\,
        b_{{\rm had}_1}\!(x,Q) \,
        \bar{b}_{{\rm had}_2}\!\!\left(\frac{\tau}{x},Q\!\right)
        +(b\leftrightarrow \bar{b}) \right]  \,,
\label{eq:cxhadronic.diff}
\end{equation}
where $\tau \equiv \hat{s}/s$, with $s$ the centre-of-mass
energy squared of the considered hadron collider. The symbols 
$b_{{\rm had}_i}(x,Q)$ and $\bar{b}_{{\rm had}_i}(x,Q)$ indicate 
the $b$- and $\bar{b}$-quark distribution functions in the 
hadron ${\rm had}_i$, and had$_1$ had$_2$ are $pp$ at the LHC 
(they would be $p \bar{p}$ at the Tevatron).  
The partonic cross section 
$\hat{\sigma}^{f.s.}(\hat{s})$ can be written in a compact
way for both final states $\bar{\mu} \mu$ and $\gamma \gamma$ 
as follows:
\begin{equation}
\hat{\sigma}^{f.s.}(\hat{s})=
\frac{\ g_b^2 \, g_{f.s.}^2 \ }{16\pi \hat{s}}
\frac{\ \beta_{f.s.}\ }{\beta_b}
\ \frac{1}{4}\ \frac{1}{3} \ \frac{1}{N_{f.s.}}
\sum_{\sigma\,,\lambda}
\left|\langle\lambda;\sigma\rangle^{f.s.}\right|^2 \,,
\label{eq:cxparton.fs}
\end{equation}
where $\beta_f=({1-4m_f^2/\hat{s}})^{1/2}$, 
$g_f=g\, m_f/2 M_W = m_f/v$, and 
\begin{equation}
g_{f.s.} = \left\{
 \begin{array}{l} 
 \alpha \sqrt{\hat{s}}/{4 \pi v} \\  g_\mu  \end{array} \right. 
\hspace{0.5truecm} 
\beta_{f.s.} = \left\{
 \begin{array}{l}  1 \\  \beta_\mu  \end{array} \right.
\hspace{0.5truecm} 
N_{f.s.} = \left\{ 
 \begin{array}{l}  2 \\ 1  \end{array} \right.
\hspace{0.5truecm} 
{\rm for} \ \ 
 f.s. = \left\{
 \begin{array}{l} \gamma \gamma \\ \bar{\mu} \mu  \end{array} \right. 
\,.
\label{eq:symbols}
\end{equation}
Finally, $\langle\lambda;\sigma\rangle^{f.s.}$ is the reduced 
helicity amplitude for the process $\bar{b} b \to H_i \to f.s.$ and 
$\sum_{\sigma\,,\lambda}$ indicate the sum over the helicities of the 
initial $b$-quarks, $\sigma$, and of the outgoing $\gamma$'s or 
$\mu$'s, $\lambda$. For $f.s. = \gamma \gamma$ it is:
\begin{equation}
\langle\lambda;\sigma\rangle^{\gamma\gamma}
\equiv
\sum_{i,j}
\left(\sigma\beta_b
       g^S_{H_i \bar{b}b}+i g^P_{H_i \bar{b}b} \right)\, 
D_{ij}(\hat{s})\,
\left[ S^\gamma_j(\hat{s})-i\lambda P^\gamma_j(\hat{s})\right] \,,
\label{eq:helamplit.gamma}
\end{equation}
for $f.s. = \bar{\mu} \mu$:
\begin{equation}
\langle\lambda ;\sigma\rangle^{\mu\mu}
\equiv
\sum_{i,j}
\left(\sigma\,\beta_b\, 
          g^S_{H_i \bar{b}b} +i\,g^P_{H_i\bar{b}b}\right)\,
D_{ij}(\hat{s})\,
\left(\lambda\,\beta_{\mu}\, 
          g^S_{H_j\bar{\mu}\mu} -i\,g^P_{H_j\bar{\mu}\mu}\right)\,.
\label{eq:helamplit.muon}
\end{equation}
In both, $D_{ij}$ is the $3 \times 3$ propagator matrix, which, as
already mentioned, has negligible off-diagonal terms in the 
specific case under consideration;
$g^{S,P}_{H_i \bar{b}b}$ are the 
couplings denoted as $g^{S,P}_{H_i}$ in Eqs.~(15) and~(16) of
Ref.~\cite{Borzumati:2004rd}, and the symbols 
$g^{S,P}_{H_j\bar{\mu}\mu}$ are:
$g^{S}_{H_j\bar{\mu}\mu} = O_{\phi_1 j}/\cos\beta$ and
$g^{P}_{H_j\bar{\mu}\mu} = -O_{a j} \tan \beta$.
The effective neutral Higgs boson couplings to two photons 
$S^\gamma_j(\hat{s})$ and $P^\gamma_j(\hat{s})$ can be found 
in Ref.~\cite{Lee:2003nt}.

As previously stated, in our specific case, 
near $\sqrt{\hat{s}}=m_{H_i}$, the cross section  
$\hat{\sigma}^{f.s.}$ is well approximated 
by the cross section with a single $\hat{s}$-channel resonance 
with mass $m_{H_i}$ and width $\Gamma_{H_i}$. The form of the 
helicity amplitudes in Eqs.~(\ref{eq:helamplit.gamma}) 
and~(\ref{eq:helamplit.muon}) can be simplified in such a 
way that $\hat{\sigma}^{f.s.}$ and 
${\rm d}\sigma^{f.s.}/{\rm d} \sqrt{\hat{s}}$ 
reduce to the simple forms:
\begin{eqnarray}
\hat{\sigma}^{f.s.}(\hat{s}) 
&  \stackrel{\approx}{_{\hat{s}\to m_{H_i}^2}}  & 
\frac{1}{\pi} \,
\frac{\sigma(\bar{b} b \to H_i)}{\Gamma_{H_i}} 
\, BR(H_i\to f.s.)\, m_{H_i} \,,
\\
\frac{{\rm d}\sigma^{f.s.}}{{\rm d} \sqrt{\hat{s}}} 
&  \stackrel{\approx}{_{\hat{s}\to m_{H_i}^2}}  &
\frac{2}{\pi} \,
\frac{\sigma(\bar{b} b \to H_i)}{\Gamma_{H_i}} 
\, BR_(H_i\to f.s.) 
\left(\!\tau\frac{\rm{d}{\cal L}^{\bar{b}b}}{{\rm d}\tau}\!\right)
\,,
\label{eq:cx.approximated}
\end{eqnarray}
where $\sigma(\bar{b} b \to H_i)$ is the partonic cross
section for the production of $H_i$ through $b$-quark fusion, 
shown in Ref.~\cite{Borzumati:2004rd}.
The dominant contribution to the widths $\Gamma_{H_1}$ and 
$\Gamma_{H_2}$ comes from the decays $H_{1,2} \to \bar{b} b$, 
at $\Phi_{A\mu} \sim 100^\circ$. In this region, 
therefore, the ratios 
$\sigma(\bar{b} b \to H_i)/\Gamma_{H_i}$ are roughly 
independent of $\Phi_{A\mu}$, whereas   
$\sigma(\bar{b} b \to H_i)$ and $\Gamma_{H_i}$, 
separately, are strongly dependent on it. 
Thus, still for $\Phi_{A\mu} \sim 100^\circ$, 
given the degeneracy of $H_1$ and $H_2$, 
the relative heights of the peaks of
$\hat{\sigma}^{f.s.}$ and
${\rm d}\sigma^{f.s.}/{\rm d} \sqrt{\hat{s}}$ 
at $\hat{s}=m_{H_{1}}^2$ and $\hat{s}=m_{H_{2}}^2$  
are practically determined by $BR(H_i\to f.s.)$ only.

\begin{figure}[p]
\begin{center}
\includegraphics[width=5.1cm]{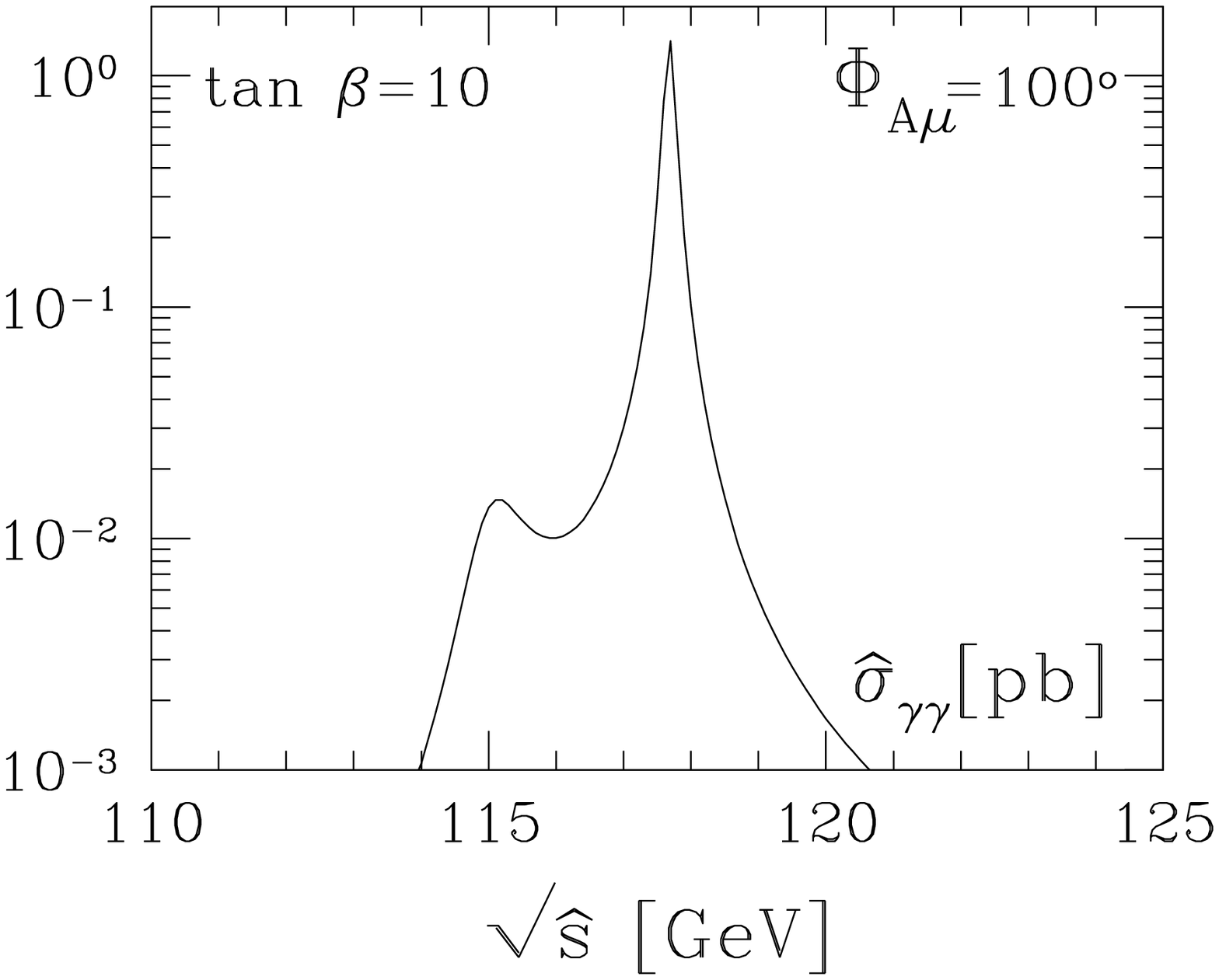}
\hspace{1mm}
\includegraphics[width=5.1cm]{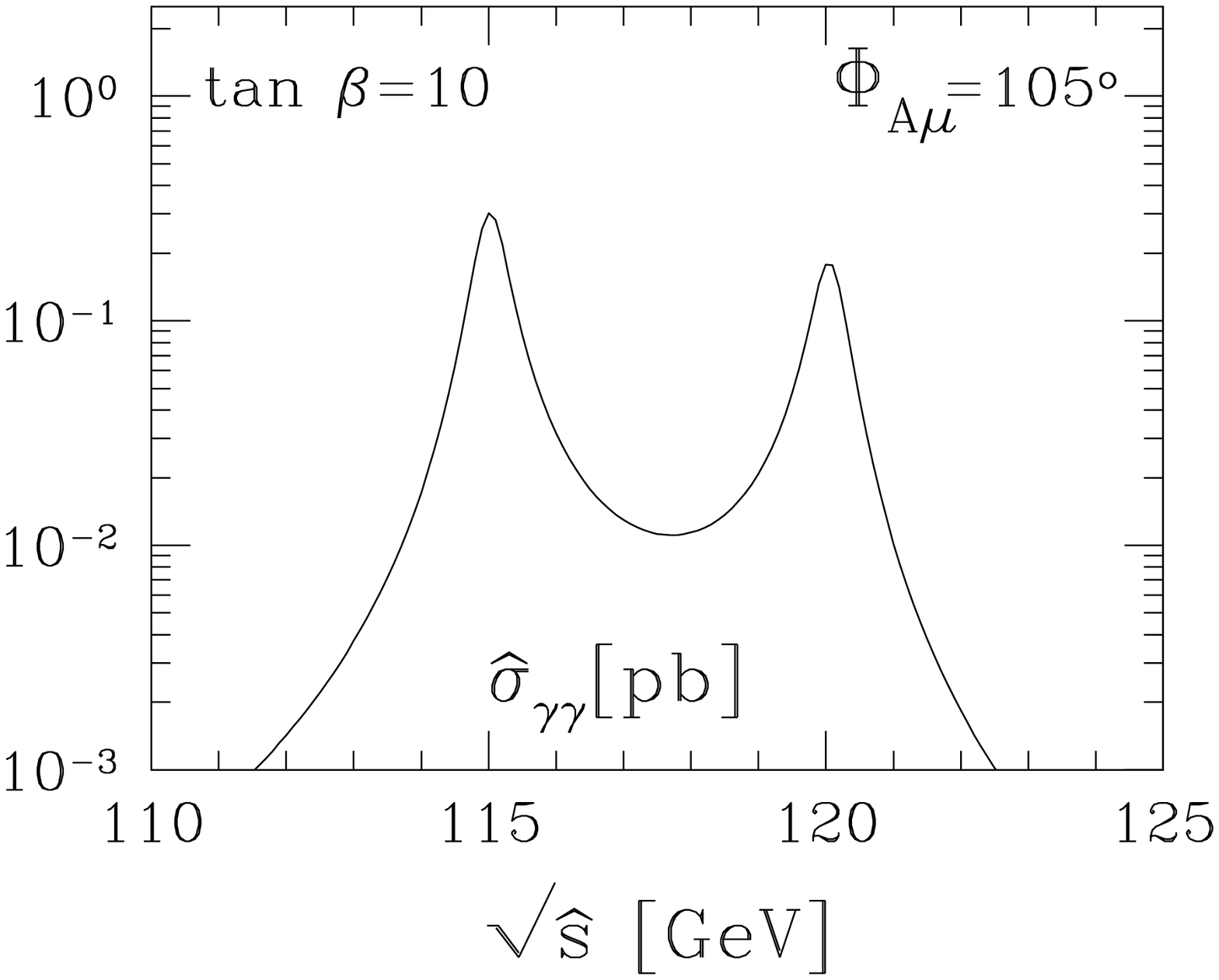}
\end{center}
\vspace{-0.5cm}
\begin{center}
\includegraphics[width=5.1cm]{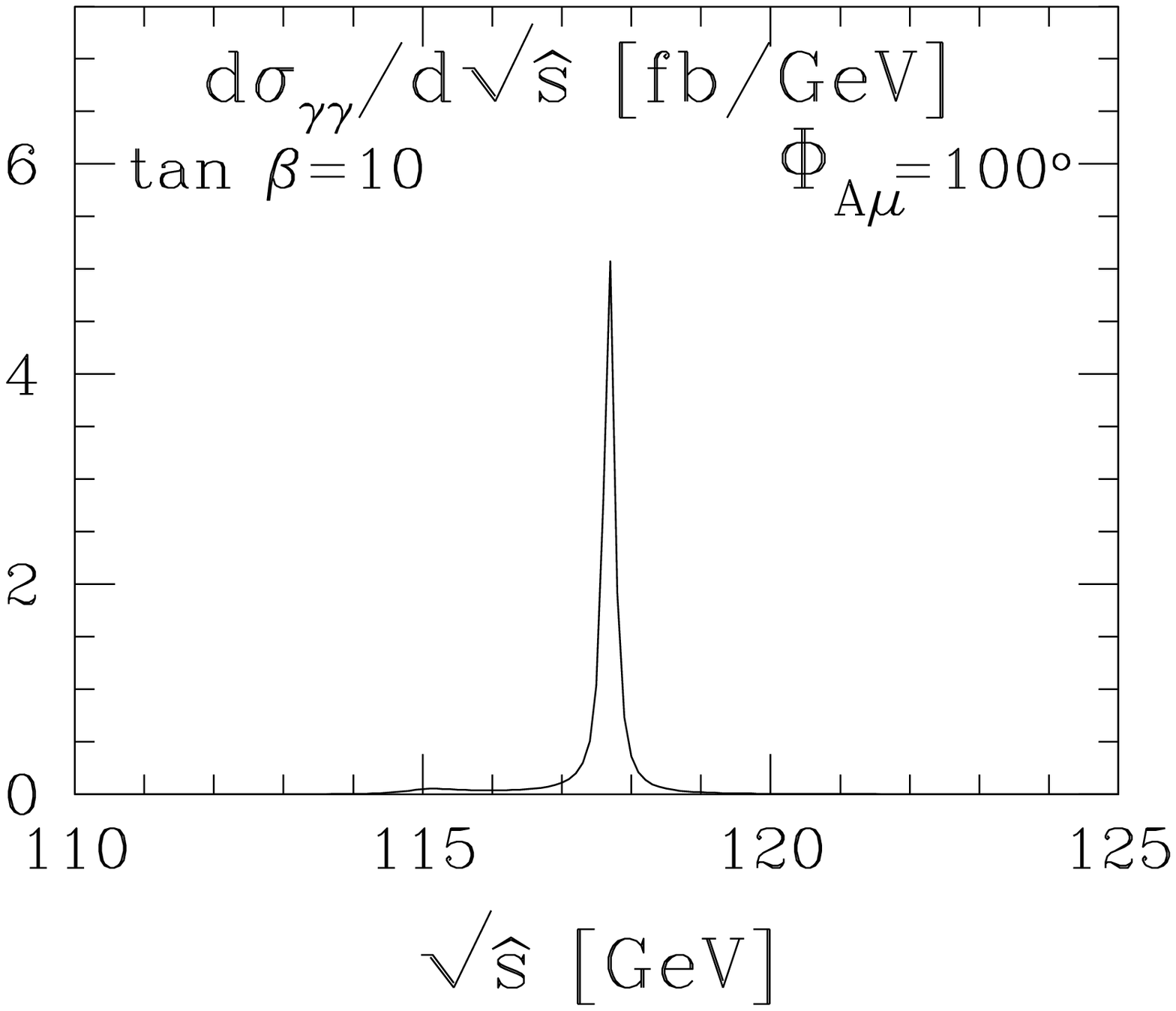}
\hspace{1mm}
\includegraphics[width=5.1cm]{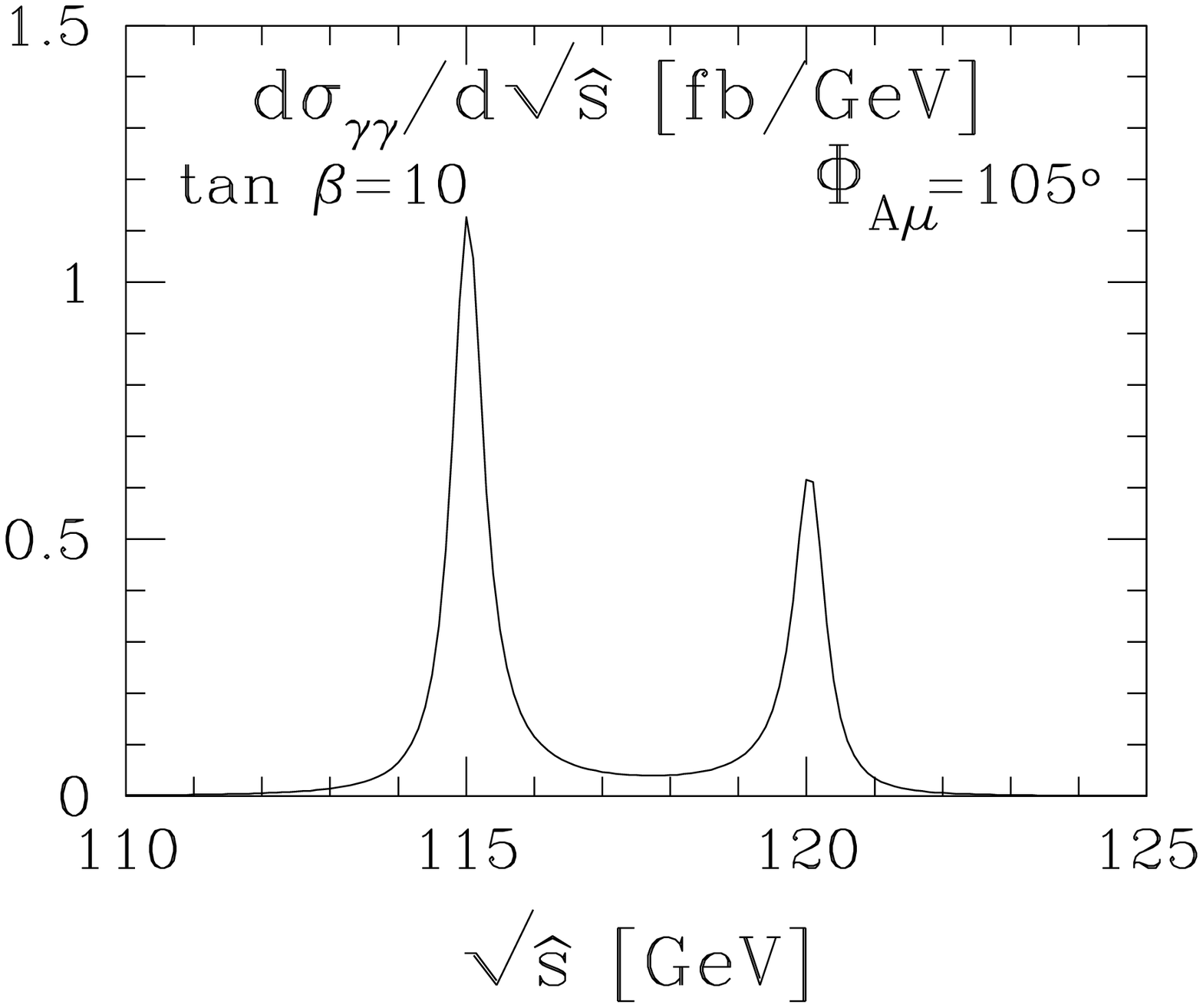}
\end{center}
\vspace{-0.5cm}
\caption{{\small \it  Cross sections
 $\hat{\sigma}^{\gamma \gamma}$  (upper two frames)
 ${\rm d}\sigma^{\gamma \gamma}/{\rm d} \sqrt{\hat{s}}$
 (lower two frames)
 at $\Phi_{A\mu} = 100^{\circ}$ and $\Phi_{A\mu} = 105^{\circ}$,
 versus $\sqrt{\hat{s}}$. In all frames, it is 
 $\tan \beta = 10$ and  $\Phi_{g\mu} = 180^{\circ}$.}}
\label{fig:cross_gamgam}
%
\vspace{0.3cm}
\begin{center}
\includegraphics[width=5.1cm]{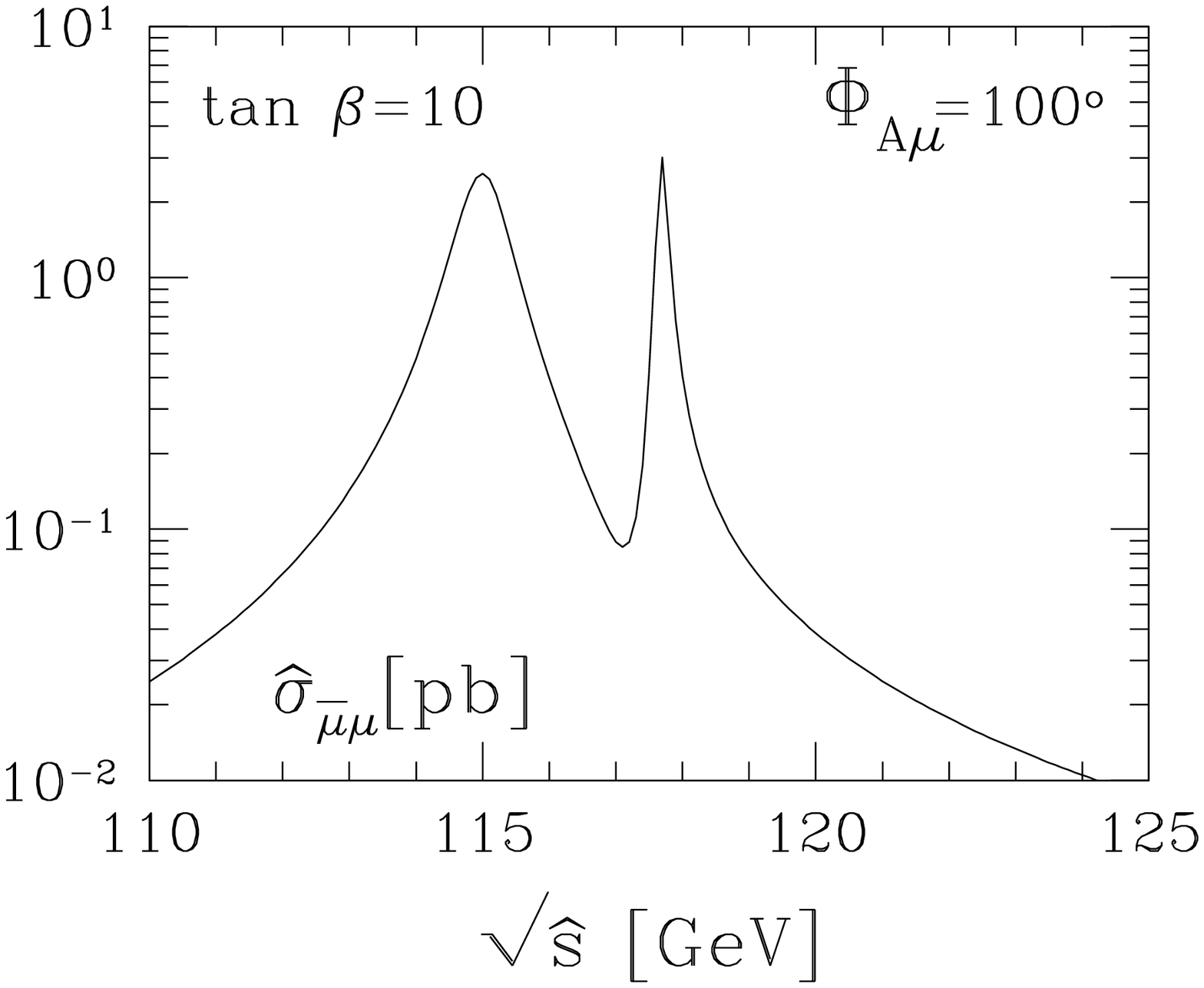}
\hspace{1mm}
\includegraphics[width=5.1cm]{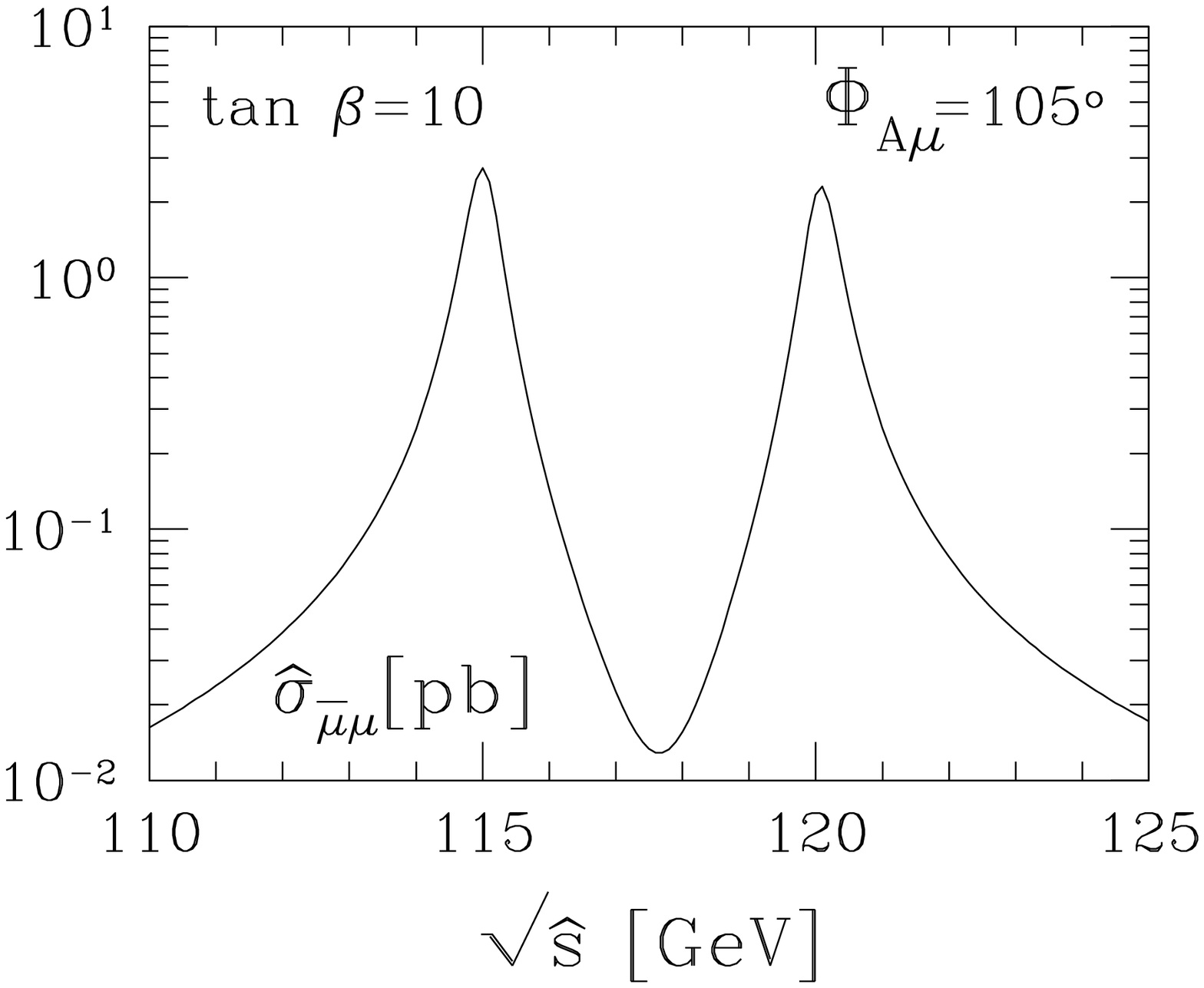}
\end{center}
\vspace{-0.5cm}
\begin{center}
\includegraphics[width=5.1cm]{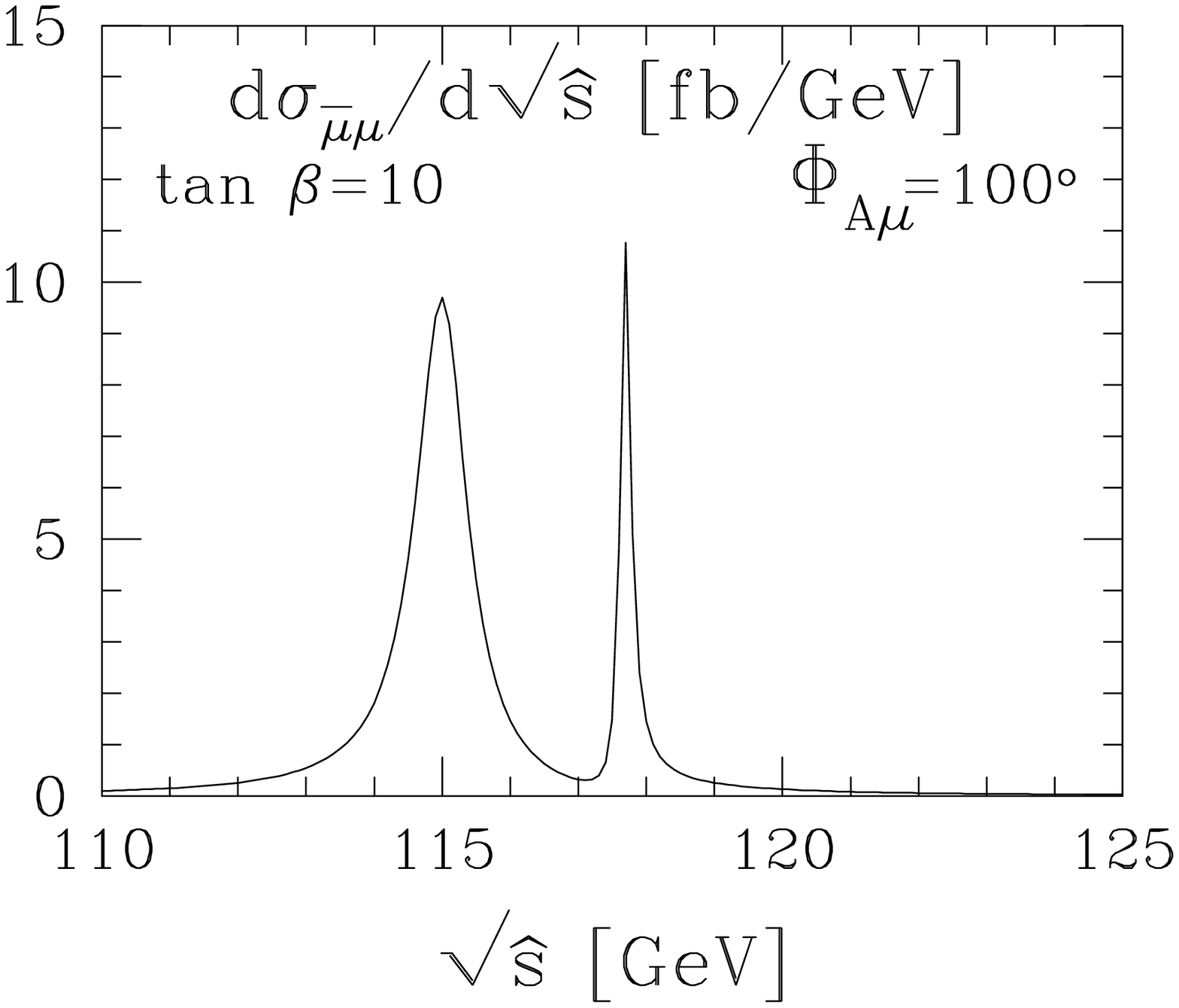}
\hspace{1mm}
\includegraphics[width=5.1cm]{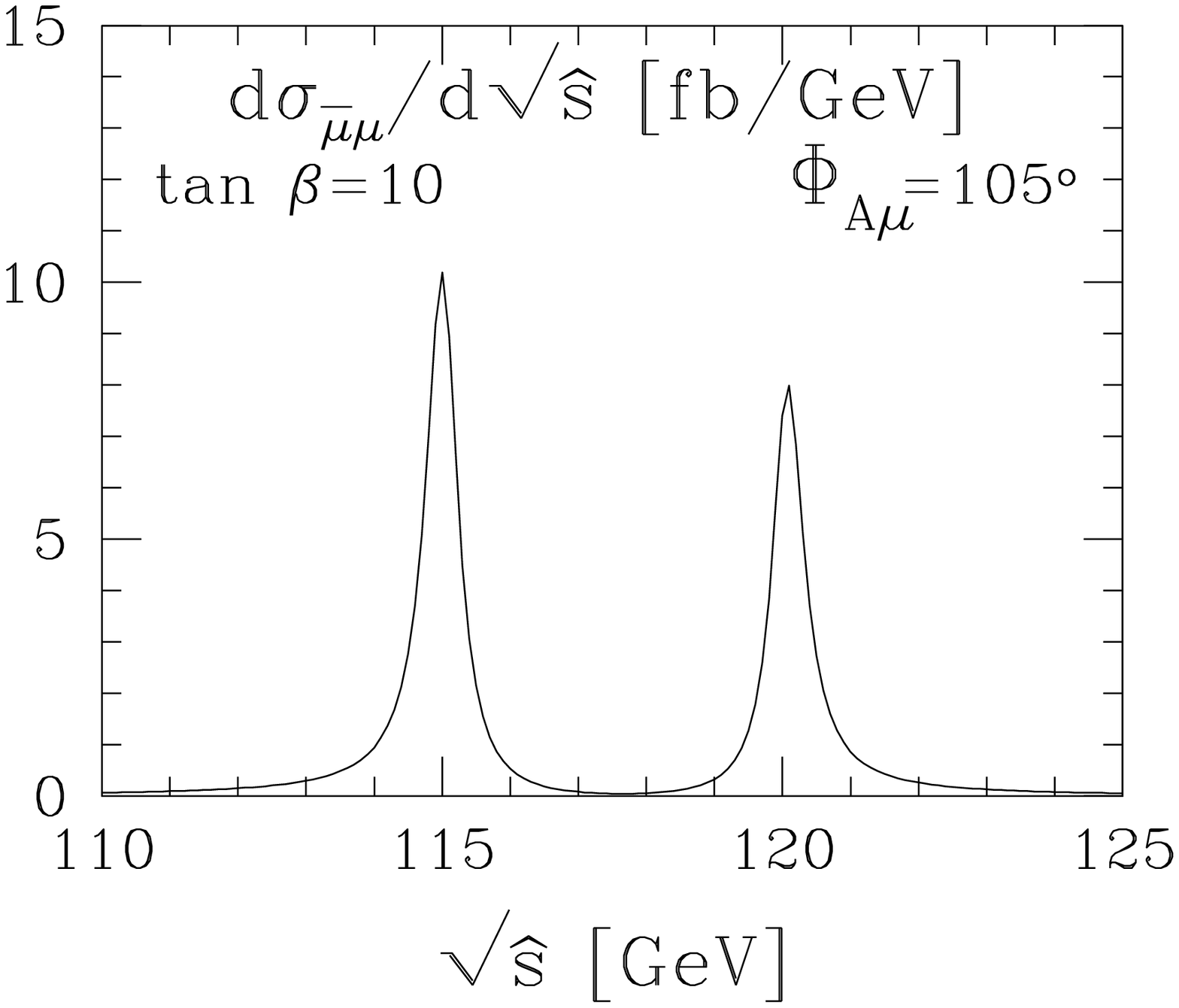}
\end{center}
\vspace{-0.5cm}
\caption{{\small \it Same as in Fig.~\ref{fig:cross_gamgam}
 for $\hat{\sigma}^{\bar{\mu} \mu}$ (upper two frames)
 and ${\rm d}\sigma^{\bar{\mu} \mu}/{\rm d} \sqrt{\hat{s}}$
 (lower two frames).}}
\label{fig:cross_mumu}
\end{figure}

In Figs.~\ref{fig:cross_gamgam} and~\ref{fig:cross_mumu}, we 
compare the values of both, the partonic cross section 
$\hat{\sigma}^{f.s.}$ and the differential hadronic cross section 
${\rm d}\sigma^{f.s.}/{\rm d} \sqrt{\hat{s}}$ for 
$f.s.= \gamma \gamma$ and $f.s.=\bar{\mu} \mu$ at two different
values of $\Phi_{A\mu}$: $100^\circ$ and $105^\circ$. These two 
values of $\Phi_{A\mu}$ are sufficiently close to avoid a 
substantial reduction of the enhancing factor for 
$\sigma(\bar{b} b \to H_1)$ 
when going from $\Phi_{A\mu}= 100^\circ$ 
to $\Phi_{A\mu}=105^\circ$. They are, however, separated enough 
for us to escape at $\Phi_{A\mu}=105^\circ$ the strong suppression 
that $\sigma(\bar{b} b \to H_2)$ and $\Gamma_{H_2}$ 
have at $\Phi_{A\mu}=100^\circ$ .

\begin{table}[t]
\caption[]{\it 
For $H_1$ and $H_2$ at two different values of $\Phi_{A\mu}$: 
$100^\circ$, $105^\circ$, we list here the values of masses and 
widths (in GeV), of the partonic cross sections
for their resonant production (in pb) and of the branching ratios 
of their decays into a pair of $\mu$'s and a pair of $\gamma$'s. 
$\Phi_{g\mu}$ is fixed at $180^\circ$ and $\tan \beta$ at 10.}
\renewcommand{\arraystretch}{1.5}
\begin{center}
\begin{tabular}{|c|c||c|c|c|c|c|} \hline
                                              & 
$\, \Phi_{A\mu}                        \, $ &
$ m_{H_i}\,[{\rm GeV}]                    $ &  
$ \Gamma_{H_i}\,[{\rm GeV}]               $ &
$ \sigma(\bar{b} b \to H_i)\,[{\rm pb}]   $ &  
$ BR(H_i\to\bar \mu \mu)                  $ & 
$ BR(H_i\to\gamma \gamma)                 $ \\
\hline\hline 
 \ $H_1$ \ & $100^{\circ}$ & $115.0$ &  0.9157   & 786.7
           &  
$8.227\times 10^{-5}$   & $3.996\times 10^{-7}$ \\ 
\hline
 \ $H_1$ \  & $105^{\circ}$ & $115.0$ &  0.5664  & 486.3
            & 
 $8.654\times 10^{-5}$   & $9.883\times 10^{-6}$ \\ 
\hline
 \ $H_2$ \  & $100^{\circ}$ & $117.7$ &  0.1818  & 145.3
            & 
  $9.978\times 10^{-5}$   & $4.858\times 10^{-5}$ \\ 
\hline
 \ $H_2$ \  & $105^{\circ}$ & $120.1$ &  0.5652  & 429.4
            & 
 $7.953\times 10^{-5}$   & $6.219\times 10^{-6}$ \\ 
\hline
\end{tabular}
\end{center}
\label{tab:mass_width_BRs}
\end{table}

For an easier comparison of the cross sections obtained for the
two different decay channels of $H_1$ and $H_2$, we list 
explicitly in Table~\ref{tab:mass_width_BRs} the values of 
$m_{H_i}$, $\Gamma_{H_i}$, $\sigma(\bar{b} b \to H_i)$, 
$BR(H_i\to \gamma \gamma)$, and $BR(H_i\to\bar{\mu} \mu)$ 
at $\Phi_{A\mu}=100^\circ$ and $105^\circ$.
We notice:
\begin{itemize}
\item
Going from $\Phi_{A\mu}=100^\circ$ to 
$\Phi_{A\mu}=105^\circ$,
the mass of $H_2$ increases by less than $2.5\,$GeV. It
is, at these two values of $\Phi_{A\mu}$, respectively
only about $3$ and $5\,$GeV larger than 
$m_{H_1}$, which, as 
explained in Ref.~\cite{Borzumati:2004rd}, 
has been fixed to $115\,$GeV for all values of $\Phi_{A\mu}$. 
\item
At $\Phi_{A\mu}=100^\circ$, as already observed in   
Ref.~\cite{Borzumati:2004rd}, the value of the 
partonic cross sections
$\sigma(\bar{b} b \to H_1)$ is about five times larger 
than $\sigma(\bar{b} b \to H_2)$. Similarly, $\Gamma_{H_2}$ 
is suppressed with respect to $\Gamma_{H_1}$ also by a 
factor of five. At $\Phi_{A\mu}=105^\circ$, on the contrary,
both cross sections and widths for 
$H_1$ and $H_2$ are remarkably similar, and only a factor 
of 1.5-2 less than the maximal values of $\sigma(\bar{b} b \to H_1)$ 
and $\Gamma_{H_1}$ obtained at $\Phi_{A\mu}=100^\circ$.  
\item
As for the branching ratios of $H_1$ and $H_2$ into 
$\gamma \gamma$ and $\bar{\mu} \mu$, we notice that at  
$\Phi_{A\mu}=100^\circ$, the branching ratio
$BR(H_1\to \gamma\gamma)$ is about two order of magnitude 
smaller than $BR(H_2\to \gamma\gamma)$, whereas it is of the 
same order of (actually 50\% larger than)
$BR(H_2\to \gamma\gamma)$ at $\Phi_{A\mu}=105^\circ$. 
The strong suppression of $BR(H_1\to \gamma\gamma)$ at
$\Phi_{A\mu}=100^\circ$ is due to the large component of
$a$ in $H_1$ at this value of $\Phi_{A\mu}$.
On the contrary, the branching ratios 
$BR(H_1\to \mu\mu)$ and $BR(H_2\to \mu\mu)$   
are of the same size for both values of $\Phi_{A\mu}$ 
considered.
\end{itemize}

Therefore, at $\Phi_{A\mu}=100^\circ$,
when the neutral Higgs bosons decay into a pair of 
photons, we expect to see only one peak corresponding to $H_2$.
Given the values of the invariant mass distribution in the 
lower-left frame of Fig.~\ref{fig:cross_gamgam},
for a luminosity of $100\,$fb$^{-1}$, we expect to have
$\sim 50$ events in the $\sqrt{\hat{s}}$ interval 
$[m_{H_2} - \delta M_{\gamma \gamma}/2, 
  m_{H_2} + \delta M_{\gamma \gamma}/2]$, with 
$\delta M_{\gamma \gamma}$ the experimental invariant
mass resolution mentioned above, $\sim 1\,$GeV.  
When $\Phi_{A\mu}=105^\circ$, it is possible to detect
two peaks and have, for the same luminosity, 
more than 30 (20) events in the two 
$\sqrt{\hat{s}}$ intervals of $1\,$GeV centered  
around
$m_{H_1}$ and $m_{H_2}$.

When $H_1$ and $H_2$ decay into a muon pair, 
although the cross sections are larger, a resolution of the
two picks will be more difficult 
because of the worse experimental resolution
$\delta M_{\mu\mu}\sim 3\,$GeV.
At $\Phi_{A\mu}=100^\circ$, again for a luminosity of 
$100\,{\rm fb}^{-1}$, it is possible to have 
more than 1,000 events in the interval
$[m_{H_1} - \delta M_{\mu \mu}/2, 
  m_{H_1} + \delta M_{\mu \mu}/2]$,
and 200 events in 
$[m_{H_2} - \delta M_{\mu \mu}/2, 
  m_{H_2} + \delta M_{\mu \mu}/2]$.
Notice that ${H_2}$ is only $2.7\,$GeV away 
from $H_1$.
For $\Phi_{A\mu}=105^\circ$,
more than 300 events are expected
for both peaks, separated by $5\,$GeV, see the lower-right
frame of Fig.~\ref{fig:cross_mumu}.

For both values of $\Phi_{A\mu}$, 
by combining the muon-decay mode with the
photon-decay mode, $H_2$ can be located more precisely
and disentangled from $H_1$. At  
$\Phi_{A\mu}=105^\circ$, actually, two well separated 
peaks may be observed.
It is clear that these considerations are only a first step 
towards more dedicated analyses, which obviously require 
detector simulations and background studies.

\subsection*{Acknowledgements}

The authors thank G. Polesello for discussions. 
The work of JSL was supported in part by Korea Research Foundation
and the Korean Federation of Science and Technology Societies Grant
funded by Korea Government (MOEHRD, Basic Research Promotion Fund).

{\small

}
\end{document}